# IRRELEVANT SPEECH EFFECT IN OPEN PLAN OFFICES: A LABORATORY STUDY

Krist Kostallari

*INRS – National Institute of Research and Security, Vandœuvre-lès-Nancy, France*
*Univ Lyon, INSA-Lyon, Laboratoire Vibrations Acoustique, F-69621 Villeurbanne, France*
*email: krist.kostallari@inrs.fr*

Etienne Parizet

*Univ Lyon, INSA-Lyon, Laboratoire Vibrations Acoustique, F-69621 Villeurbanne, France*

Patrick Chevret, Jean-Noël Amato

*INRS – National Institute of Research and Security, Vandœuvre-lès-Nancy, France*

Edith Galy

*LAPCOS – Laboratory of Anthropology, Clinical, Cognitive and Social Psychology, Nice, France.*

It seems now accepted that speech noise in open plan offices is the main source of discomfort for employees. This work follows a series of studies conducted at INRS France and INSA Lyon based on Hongisto's theoretical model (2005) linking the Decrease in Performance (DP) and the Speech Transmission Index (STI). This model predicts that for STI values between 0.7 and 1, which means a speech signal close to 100% of intelligibility, the DP remains constant at about 7%. The experiment that we carried out aimed to gather more information about the relation between DP and STI, varying the STI value up to 0.9. Fifty-five subjects between 25-59 years old participated in the experiment. First, some psychological parameters were observed in order to better characterize the inter-subjects variability. Then, subjects performed a Working-Memory (WM) task in silence and in four different sound conditions (STI from 0.25 to 0.9). This task was customized by an initial measure of mnemonic span so that two different cognitive loads (low/high) were equally defined for each subject around their span value. Subjects also subjectively evaluated their mental load and discomfort at the end of each WM task, for each noise condition. Results show a significant effect of the STI on the DP, the mental load and the discomfort. Furthermore, a significant correlation was found between the age of subjects and their performance during the WM task. This result was confirmed by a cluster analysis that enabled us to separate the subjects on two different groups, one group of younger and more efficient subjects and one group of older and less efficient subjects. General results did not show any increase of DP for the highest STI values, so the "plateau" hypothesis of Hongisto's model cannot be rejected on the basis of this experiment.

## 1. Introduction

Noise in open plan offices is one of the most annoying sources during a workday (Boyce [1], Klitzman and Stellman [2]). In a field study, Sundstrom et al. [3] showed that 54% of interviewed employees were more annoyed by speech and telephones ringing than other types of sound. Later on, during a study from Pierrette et al. [4], 58% of 237 employees confirmed that noise was the greatest source of annoyance in open plan offices. Perrin-Jegen and Chevret [5] conducted a second survey in 23 open-plan offices, making it possible to collect the responses from 617 employees. Most of the results of the first survey were confirmed, with an increase in the significance of the statistical analyses. The principal results Pierrette et al. showed that intelligible speech caused more disruption during their everyday tasks than any other source of noise.

On the other hand, since we are interested in the impact of noise on a person, this requires some investigation into his/her psychological state and performance. This means that we should take account of which part of his cognitive process is impaired by noise. Lashey [6] and Rosenbaum et al. [7] show that memory is one of the most solicited processes during our complex behavior. More precisely, for employees in open plan offices, working-memory is the most requested of cognitive process during a workday. This process is often disrupted by noise. The effect of noise on the working-memory is called Irrelevant Sound Effect (ISE) [8]. ISE is usually evaluated by measuring the performance during a working-memory task or short-term memory task, in the presence of noise. Different laboratory studies, such as the field studies, show that intelligible speech is the most disruptive noise during a serial recall task [9, 10]. The main goal of this study is to measure the ISE as a function of speech intelligibility. In the next subsection are presented previous experiments based on the same goal but with different measurement protocols. Next, this paper will present our own experimental method that aimed to evaluate the relation between ISE and intelligibility in the light of Hongisto's model [11].

### 1.1 Review of laboratory experiments

The studies listed below are devoted to the understanding of the impact of speech intelligibility on performance during a serial-recall task. Speech intelligibility is estimated by the standardized indicator Speech Transmission Index (STI). When the intelligibility of speech is high, represented by high values of STI, it impairs the performance during the task. This indicator varies from 0 to 1, with 0 representing a low intelligible speech and 1 representing a high intelligible speech.

Ellermeier and Hellbrück [12] were the first to conduct experiments of ISE related directly to speech intelligibility. In their experiments, the STI took values from 0.2 to 1. The results showed that the performance of subjects was significantly impaired by the variation of STI. Schlittmeier et al. [13] confirmed these results, with STI values from 0.3 to 0.8. For each condition, the disturbance due to noise was also subjectively evaluated by asking participants: 'How disturbing were the background sounds in this experimental block to you?' Results showed that participants were more disturbed by noise when the STI was high. Haka et al. [14] tested the participants performance when the STI varies from 0.1 to 0.65. Their statistical analyses confirmed that this variation of STI decreased the participants' performance. Once again, high STI values lead to a high noise annoyance. Another interesting result in Haka et al. [14] is that subjective evaluation was disturbed more easily than performance. On their side, Jahncke et al. [15] used a serial-recall task of words. In their experiment, the STI varied from 0 to 0.7. They concluded that there existed a global effect of STI on the decrease of performance, but there was no significant difference between the conditions of STI = 0.35 and STI = 0.7.

At INRS and INSA Lyon, two major studies were conducted to measure the ISE in the presence of intelligible speech. The first one [16] tested the effect of intelligibility on the decrease in performance during a serial-recall task with STI = 0.25; 0.35; 0.45; 0.65. Ebissou et al. [16] also measured the workload during the experiment using the questionnaire NASA-TLX (Raw Task Load Index). The second study of Brocolini et al. [17] measured the effect of fluctuations in noise with STIt = 0.38; 0.51; 0.56; 0.69. The definition of STIt will be explained in the next subsection. As in the previous study, Brocolini et al. measured the workload using NASA-RTLX (Raw Task Load Index) [18]. In both studies, participants could be separated in two groups according to their general performance. For the highly-performing participants, the decrease in performance was significantly lower than the one of the other group.

Ebissou et al. concluded that STI had a significant effect on the decrease in performance and the workload but only for the group of the "bad-performers". On the other side, Brocolini et al. showed that STIt had no effect on decrease in performance, or on the workload, even if the participants were separated in two groups.

### 1.2  Using STI for the measurement of the ISE

The main motivation of this research work is to study the effect of intelligible speech on performance during a working-memory task. This subsection will be separated in three parts: intelligibility indicators, ISE as a function of STI, and the motivations for a new experiment.

#### 1.2.1  Intelligibility indicators

The intelligible speech used in studies previously mentioned is represented by its STI values.

The STI is known as a good estimator of intelligibility of words and sentences signals in the presence of a stationary noise. It was initially developed by Steeneken and Houtgast [19]. This indicator is defined as:

$$STI = \sum_{k=1}^{n} w_k TI_k$$

Where the $w_k$ is a weight factor for each octave band $k$ and the $TI_k$ is the transmission index that represents the modulation losses of the signal when it is transmitted from the source to the listener's ear. This transmission index is calculated as:

$$TI_{k,i} = \frac{SNR_{k,i}^{app} + 15}{30}$$

Where $SNR_{k,i}^{app}$ is the apparent signal-to-noise ratio at the listener's ear level and is calculated in each octave band for every modulation frequency corresponding to third octave bands from 0.63 Hz to 12.5 Hz. For $SNR_{k,i}^{app} \geq +15$ a speech signal is completely intelligible and for $SNR_{k,i}^{app} \leq -15$ dB such a signal is not intelligible at all. This signal-to-noise ratio takes into account different modulation losses that the speech signal suffers when it goes from the source to the listener's ear, in a room with ambient noise considered as stationary. A non-stationary version (STIt) had been proposed by Brocolini et al. [17] to take into account temporal fluctuations of noise, which increases intelligibility ("listening in the dips" effect). It is based on the use of a sliding window just like ESII of Rherbergen [20].

#### 1.2.2  Measurement of the Irrelevant Speech effect: Hongisto's model

Using a huge number of results found in the literature, Hongisto [11] conjectured a model linking the measurement of the ISE and the STI. In this model, he presented the principle of decrease in performance that is calculated as the difference of the performance acheived in a control condition, generally silence, and the performance achieved in a sound condition:

$$DP = Perf_{control} - Perf_{noise}$$

According to Hongisto, the relation between DP and STI is a sigmoid, represented in figure 1 together with the results of the studies previously mentioned.

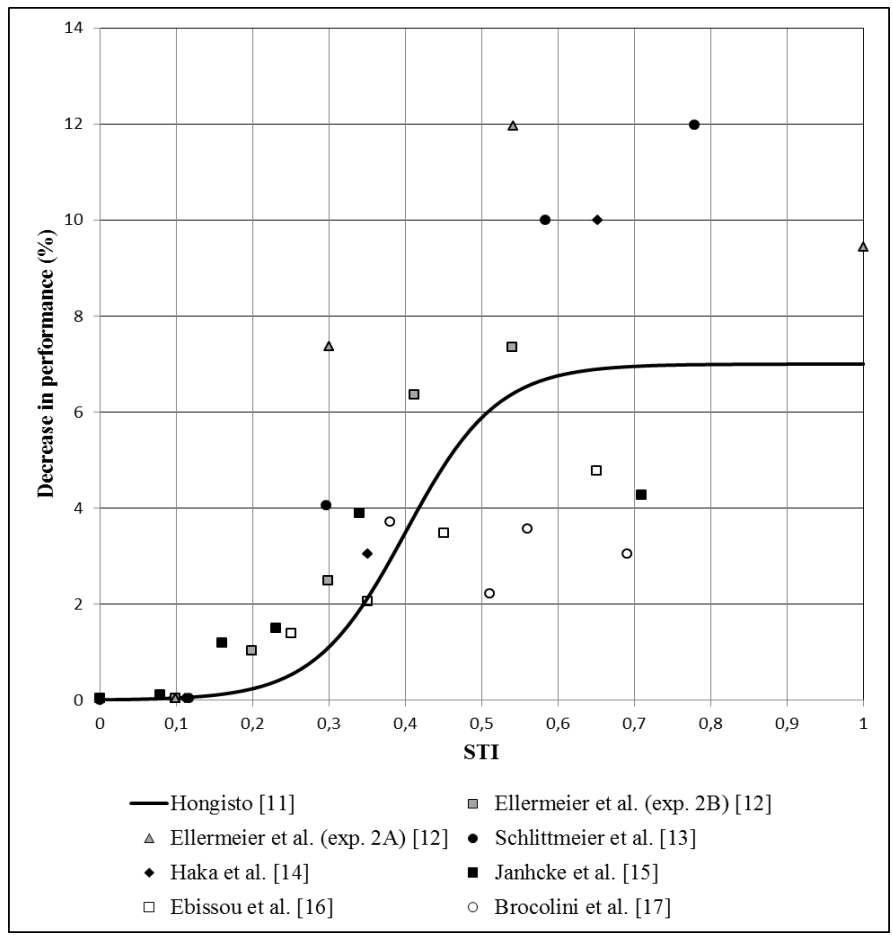

Figure 1. Decrease in performance as a function of STI.

In this figure, Hongisto's model (full line) indicates that for values of STI between 0.25 and 0.7 the DP varies significantly. For STI > 0.7 the DP remains constant. Furthermore, it is obvious from the figure that there are not a lot of experiments that investigate the comportment of DP for values of STI higher than 0.8.

The literature review leads us to the motivation for undertaking this research, which is to complete the experimental data for the measure of the ISE (figure 1), especially in the high STI domain. Furthermore, the protocols of the studies mentioned before are different, so the interest of this study is to propose a protocol that can control a maximum of psychological and acoustical parameters. As we have showed previously in [16] and [17], workload had different results. This difference justifies the necessity of better control on psychological parameters. This is provided by measuring the psychological state through Levenson's questionnaire [21], Thayer's questionnaire [22], NASA-RTLX [18] and noise sensitivity. Moreover, STROOP task [23], and all this control will give us to a better understanding of the evolution in performance during the working memory task with high STI value.

## 2. Methods and materials

### 2.1 Experimental facilities

The experiment took place in the soundproof testing room of the Vibrations and Acoustics Laboratory (LVA) at INSA Lyon. Inside the room, the participant was seated on a chair. In front of him/her there was a table with a screen, a mouse which helped the participant to accomplish different psychological tests, and a microphone which recorded the answers during the serial-recall task. Most of the tasks were computerized. Behind the table there was a loudspeaker, higher than the weight of the screen so that the sound could directly reach the participant. The frequency response of the loudspeaker in the room was equalized to have a flat frequency response at the position of the participant.

## 2.2 Participants

55 people participated in this experiment (Age: 25-59 years, Median: 43, Gender: 26 females, 29 males). All participants were aware of the nature of the experiment. Their hearing threshold was measured before the experiment (Oscilla USB-300 Screening Audiomoter, 11 frequencies between 125 and 11000 Hz). According to the recommendations of the International Bureau for AudioPhonology (BIAP) [24] , 45 participants did not present hearing losses. 6 of them had hearing losses between 22 and 25 dB and 4 participants had hearing losses between 30 and 35 dB. According to BIAP, these participants have slight hearing losses. For them, "speech is well perceived for a normal voice, hardly perceived for a low and/or distant voice and most of familiar noises are perceived". For this experiment none of the participants were considered with a hearing loss that could question the results.

## 2.3 Procedure

During the experiment, participants accomplished one preliminary phase with psychological tests and the main phase with the serial-recall task. The whole experiment lasted approximately 1 hour and 30 minutes, 40 minutes for the preliminary phase and 50 minutes for the main phase.

### 2.3.1 Preliminary phase

The following tasks were submitted to the participants:

*Mnemonic span task* [25] measured the maximal capacity of a person to memorize lists with a given number of elements and to restore them right after hearing them. The task started with two elements per memorized list. Every two lists the number of elements grew by one. The task stopped when the participant made two mistakes in two consecutive lists. The maximal capacity was equal to the number of elements of the last list correctly restored.

After the measurement of the mnemonic span, all the other tasks were computerized. For each task/questionnaire, guidelines appeared on the screen before the beginning of the task/questionnaire.

*STROOP task* aimed to measure the sensibility of the participant in the presence of an interference effect. This last is defined as the psychological perturbations due to a previous learning process during a new learning process similar to the first one [23]. In this study, the 'learning process' is identified by the working memory task and the interference effect is the type of noise, that changes from one condition to the other.

*Levenson's questionnaire* measured the Locus Of Control (LOC) of a person. This questionnaire indicates if the person is easily disturbed by external disruptive factors while he/she is doing a task that requires concentration.

### 2.3.2 Main phase

The main phase was designed to measure the effect of noise in performance and annoyance during a working memory task. At this stage of the experiment, the participant repeated the procedure in five different sound conditions. The participant had to fill up in the running order three questionnaires and to complete the main task, explained in the next paragraph.

*Thayer's questionnaire* was given to participants in the beginning of the main phase and after every sound condition. This questionnaire was used to evaluate the vigilance state of the participant throughout twenty adjectives.

*Working memory task.* There were five sound conditions where the participant had to repeat the working-memory task. For every condition the participant had to memorize 16 lists of words. The number of words that appeared in each list was different. This number was defined by the mnemonic span value $n$, evaluated for each participant during the preliminary phase. 8 lists had $n+2$ words to memorize and the other 8 had $n-1$ words. Respectively, the participant went through two different cognitive loads, high and low. These loads were presented randomly for each sound condition. Since, the mnemonic span value is different from one participant to another; this procedure customizes the task for each individual.

After each sound condition the participant had to fill up three questionnaires:
- *NASA-RTLX (Raw Task Load Index)* which evaluates the workload through 5 questions on mental requirement, time requirement, the performance, the provided effort and the frustration.
- *Annoyance* was evaluated throughout one simple question: 'Were you disturbed by the background noise?'
- *Thayer's questionnaire* mentioned previously.

## 2.4   Background sounds

The participants repeated the main phase 5 times, in silence and in 4 different sound conditions. The sounds were all played at an acoustic level of 55 dB (A). Each background sound was made up by a mix of a speech signal and babble noise. The speech was taken from a database of audio books found on the internet [26]. Then, the speech signal was cut and cleaned up. Therefore, the long-term average speech spectrum (LTASS) was applied to both speech and babble noise. Next, both signals were normalized with respect to their root mean square that represents the acoustical effective power of each signal. Finally, 4 different Signal-to-Noise Ratios (SNR) were applied between the speech signal and the babble noise in order to obtain 4 types of background sounds (1 for each sound condition). The applied SNR gave 4 types of signals with STI = 0.25, 0.45, 0.75, and 0.9. Then, for each sound condition, one type of 9-minute-signal was played at a level of 55 dB (A) at the position of the participant.

## 3.   Results

We present here, the most important result, concerning Hongisto's model.

The statistical analyses were done with STATA14. During the experiment, the participant repeated the same procedure several times. Therefore, his/her performance from one condition to the other can be similar. A standard analysis of variance (ANOVA) cannot take the differences in-between subjects into account. For this purpose the appropriate analysis can be the repeated ANOVA measure. The result of decrease in performance (DP) due to sound is shown in figure 2. The repeated ANOVA measure shows a highly significant effect of the STI on the decrease in performance ($F(3) = 4.32$ $p = 0.0059$). This effect confirms the other results found in the studies mentioned previously [12-17]. More precisely, decrease of performance is constant for the two highest STI values, confirming the plateau of the curve proposed by Hongisto (see figure 1).

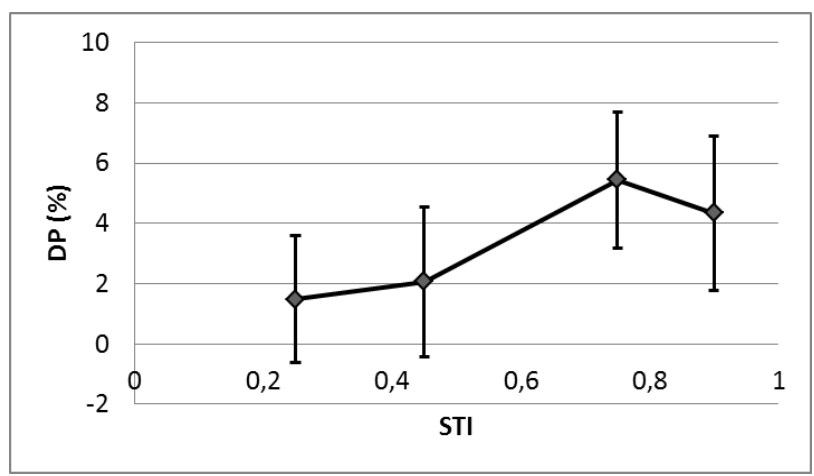

Figure 2: The decrease in performance (DP) as a function of Speech Transmission Index (STI).

The other statistical analyses are still ongoing and their results will be shown during the presentation of this study.